\documentclass[manuscript]{aastex}

\slugcomment{In review with Icarus}

\shorttitle{Latitudinal variation of methane (CH$_4$) in Neptune's atmosphere.}
\shortauthors{Irwin et al.}

\usepackage{lineno}

\begin{document}


\title{Latitudinal variation in the abundance of methane (CH$_4$) above the clouds in Neptune's atmosphere from VLT/MUSE Narrow Field Mode Observations.}


\author{Patrick G. J. Irwin, Daniel Toledo, Ashwin S. Braude}
\affil{Department of Physics, University of Oxford, Parks Rd, Oxford, OX1 3PU, UK.}
\author{Roland Bacon}
\affil{CRAL - Observatoire de Lyon, 9 Avenue Charles Andr\'e, 69230 Saint-Genis-Laval, France.}
\author{Peter M.\ Weilbacher}
\affil{Leibniz-Institut f\"ur Astrophysik Potsdam (AIP), An der Sternwarte 16, D-14482 Potsdam, Germany.}
\author{Nicholas A. Teanby}
\affil{School of Earth Sciences, University of Bristol, Wills Memorial Building, Queens Road, Bristol, BS8 1RJ, UK.}
\author{Leigh N. Fletcher}
\affil{Department of Physics \& Astronomy, University of Leicester, University Road, Leicester, LE1 7RH, UK.}
\author{Glenn S. Orton}
\affil{Jet Propulsion Laboratory, California Institute of Technology, 4800 Oak Grove Drive, Pasadena, CA 91109, USA.}
\email{patrick.irwin@physics.ox.ac.uk}



\begin{abstract}
Observations of Neptune, made in 2018 using the new Narrow Field Adaptive Optics mode of the Multi Unit Spectroscopic Explorer (MUSE) instrument at the Very Large Telescope (VLT) from 0.48 -- 0.93 $\mu$m, are analysed here to determine the latitudinal and vertical distribution of cloud opacity and methane abundance in Neptune's observable troposphere (0.1 -- $\sim3$ bar). Previous observations at these wavelengths in 2003 by HST/STIS (Karkoschka and Tomasko 2011, Icarus 205, 674--694)  found that the mole fraction of methane above the cloud tops (at $\sim 2$ bar) varied from $\sim 4$\% at equatorial latitudes to $\sim 2$\% at southern polar latitudes, by comparing the observed reflectivity at wavelengths near 825 nm controlled primarily by either methane absorption or H$_2$--H$_2$/H$_2$--He collision-induced absorption. We find a similar variation in cloud-top methane abundance in 2018, which suggests that this depletion of methane towards Neptune's pole is potentially a long-lived feature, indicative of long-term upwelling at mid-equatorial latitudes and subsidence near the poles. By analysing these MUSE observations along the central meridian with a retrieval model, we demonstrate that a broad boundary between the nominal and depleted methane abundances occurs at between 20 -- 40$^\circ$S. We also find a small depletion of methane near the equator, perhaps indicating subsidence there, and a local enhancement near $60 - 70^\circ$S, which we suggest may be associated with South Polar Features (SPFs) seen in Neptune's atmosphere at these latitudes. Finally, by the use of both a reflectivity analysis and a principal component analysis, we demonstrate that this depletion of methane towards the pole is apparent at all locations on Neptune's disc, and not just along its central meridian.

 \end{abstract}





\section{Introduction}

\begin{figure}
\epsscale{0.4}
\plotone{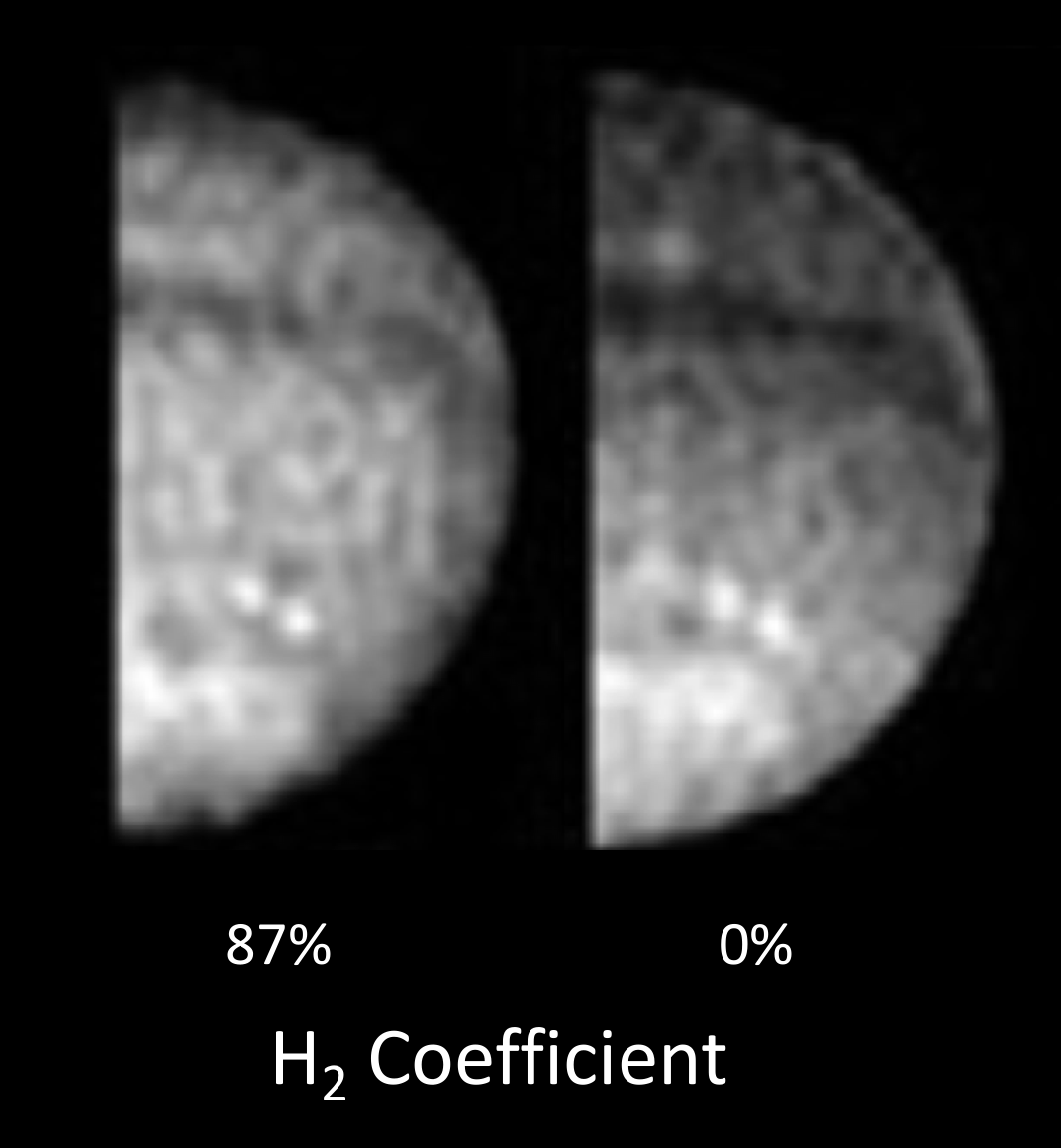}
\caption{Reproduction of part of Fig. 5 of \cite{kark11} showing HST/STIS observations of Neptune made in 2003 at two wavelengths near the 825 nm H$_2$--H$_2$/H$_2$--He CIA band, one where H$_2$--H$_2$/H$_2$--He opacity accounts for 87\% of the opacity (left) and one where the opacity is entirely from gaseous methane (right). The images here have been processed to remove discrete features \citep{kark11}. At wavelengths of low H$_2$--H$_2$ and H$_2$--He absorption  (i.e. right) we see that the polar latitudes appear to be brighter than equatorial ones. If methane abundance were assumed to be constant with latitude, the brighter appearance of polar latitudes in the methane-image could be interpreted as being due to higher clouds there. However, the more uniform brightness seen in the H$_2$--H$_2$/H$_2$--He-absorbing image (left) shows that the appearance of Neptune in the methane-absorbing images is actually due to lower methane abundance at polar latitudes. \label{stisneptune}}
\end{figure}

The visible and near-infrared spectra of both Uranus and Neptune are formed by the reflection of sunlight off its cloud layers, modulated mostly by the absorption of gaseous methane and also Rayleigh scattering at blue wavelengths. At wavelengths of low gaseous absorption, sunlight can penetrate to, and be reflected from, the deepest cloud layers, while at wavelengths of high absorption only sunlight reflected from the upper layers can be observed.  Hence, such spectra can be inverted to determine the cloud structure as a function of location and altitude, but only if we know the vertical and latitudinal distribution of methane. For many years planetary astronomers assumed that the vertical profiles of methane determined from Voyager 2 radio-occultation observations could be used at all locations on these planets. However, HST/STIS observations of Uranus recorded in 2002 \citep{kark09} and similar observations of Neptune recorded in 2003 \citep{kark11} both revealed that the abundance of methane actually varies significantly with latitude on both planets.  HST/STIS observes the 300 -- 1000 nm range, which includes collision-induced absorption (CIA) bands of H$_2$--H$_2$ and H$_2$--He near 825 nm. Such spectra allow the cloud structure to be probed at wavelengths of either mostly CH$_4$ absorption or mostly H$_2$--H$_2$/H$_2$--He collision-induced absorption, allowing variations of CH$_4$ to be differentiated from cloud-top pressure variations (Fig. \ref{stisneptune}).  \cite{kark09,kark11} found that the methane abundance near the main observable H$_2$S cloud tops at 2--3 bar varies from $\sim4$\% at equatorial latitudes to $\sim2$\% at polar latitudes for both planets. These observations point to a very different overturning circulation of air at these pressure levels than that seen in the upper troposphere, where mid-infrared observations find cooler temperatures near the tropopause at mid-latitudes than at the equator and poles, indicative of air rising at mid-latitudes in the upper troposphere and subsiding elsewhere \citep[e.g.][]{fletcher14,depater14}. These observed latitudinal dependancies of methane also mean that cloud retrievals that do not include these variations  will likely retrieve erroneously low cloud-top pressures near the poles of these planets.

Since the HST/STIS observations of Uranus in 2002 \citep{kark09}, latitudinal variation in the abundance of methane above the clouds in Uranus's atmosphere has been found in:  1) a reanalysis of HST/STIS 2002 observations \citep{srom11}; 2) IRTF/SpeX observations (800 -- 850 nm) made in 2009 \citep{tice13}; 3) Gemini/NIFS observations made in 2010 \citep{irwin12} at 1.45 -- 1.85 $\mu$m, where the degeneracy between cloud-top pressure and methane was partially broken by the use of high spectral resolution; 4) HST/STIS observations made in 2012 combined with Keck/NIRC2 observations (2007-2011) \citep{srom14}, where methane depletion at high latitudes was modelled as regions of downwelling air, using ``proportionally descended gas" profiles (found by \cite{srom11} to provide a good match to the HST/STIS 2002 observations of \cite{kark09}); and 5) further HST/STIS observations made in 2015, again combined with Keck/NIRC2 observations \citep{srom19}, which again found depleted methane towards Uranus's north pole, but found that the absolute methane abundance was dependent on the assumed aerosol properties.

While the abundance of methane in Uranus's atmosphere has been extensively studied, the latitudinal variation of methane in Neptune's atmosphere has not been determined since HST/STIS observations in 2003 \citep{kark11}. Hence, it is of great interest to see if the latitudinal variation of methane observed by HST/STIS in Neptune's atmosphere over 15 years ago remains the same today. In this paper we present new visible/near-infrared ground-based spectral observations of Neptune made in 2018 with VLT/MUSE and use them to make an initial estimation, for the first time with ground-based observations, of the spatial distribution of methane above the clouds in Neptune's atmosphere. A more exhaustive analysis of clouds and cloud-top methane abundance, combining the VLT/MUSE observations with longer wavelength near-infrared observations from the VLT/SINFONI and Gemini/NIFS will be the topic of a future, follow-on paper. 

\section{VLT/MUSE observations}
The Multi Unit Spectroscopic Explorer (MUSE) instrument \citep{bacon10} at ESO's Very Large Telescope (VLT) in Chile is an integral-field spectrograph, which records 300 $\times$ 300 pixel `cubes', where each pixel, or `spaxel', contains a complete visible/near-infrared spectrum (0.48 -- 0.93 $\mu$m) with a spectral resolving power of 2000 -- 4000. MUSE has two viewing modes: a Wide-Field Mode (WFM) with a 60\arcsec\ $\times$ 60\arcsec\ field of view (FOV) and a Narrow-Field Mode (NFM), with a  7.5\arcsec $\times$ 7.5\arcsec\ FOV (giving a pixel size of 0.025\arcsec), which utilises Adaptive Optics \citep{arsenault12,stroebele12} to improve the spatial resolution to $<  0.1$\arcsec. At such a spatial resolution it is important to correct for differential atmospheric refraction between different wavelengths, for which MUSE utilises an atmospheric dispersion compensator that reduces residual shifts between wavelengths to less than a pixel. Commissioning observations of Neptune were made in this new NFM mode on June 19th 2018, which we present and analyse here. The NFM mode needs to reserve some wavelengths for its laser guide star and thus MUSE/NFM observations do not include the wavelength range 0.578 -- 0.605 $\mu$m. Six observations of Neptune were made, two with exposures of 10s each and four with exposures of 120s, as summarised in Table \ref{tbl-1}. The performance of the adaptive optics system is estimated for these observations to achieve a point-spread-function (PSF) with a full-width-half-maximum (FWHM) of 0.06\arcsec\ at the wavelengths considered here. No attempt was made to quantify the shape of the PSF from these observations (which would require the analysis of point object observations, which we did not have) and hence it is not known how significant a fraction of the total observed flux comes from outside the 0.06\arcsec \ FWHM. Hence, we avoided analysing observations made too close to the planet's limb, which could contain an unquantified contribution from space, and observed no biases at the extremes of the latitude range considered.

In contrast to the MUSE spectral resolution of 2.45\AA\ in NFM, the best available source of methane absorption in this spectral range are the band-model coefficients of \cite{kark10}, to which we have fitted k-distributions using exponential-sum fitting. These data have a spectral resolution of 25 cm$^{-1}$ between 19300 and 25000 cm$^{-1}$ (0.518 - 0.4 $\mu$m), and 10 cm$^{-1}$ at wavenumbers less than 19300 cm$^{-1}$ (0.518 $\mu$m). These resolutions equate to roughly 5 \AA\ (i.e. 0.0005 $\mu$m) across the visible range and so the MUSE data cannot be analysed using these coefficients at their native resolution.  Hence, the data were first smoothed to make them more compatible with the available methane data and also improve the signal-to-noise (SNR) ratio. We chose to smooth the MUSE data to the resolution of the IRTF/SpeX instrument, which has a triangular instrument function with FWHM = 20 \AA\ (i.e 0.002 $\mu$m), sampled at 0.001 $\mu$m. We find that this spectral resolution captures the essential shape of the observed spectra very well and has significantly better SNR than MUSE's native resolution, giving SNRs ranging from $\sim 80$ at the reflectance peak to $\sim 40$ at methane-absorbing wavelengths.

The MUSE data were reduced with the VLT/MUSE pipeline v2.5.1 \citep{weilbacher14} and flux-calibrated using the spectrophotometric standard star EG274, observed on the same night. For this study, spectra were extracted from the MUSE data from 64 individual spectral pixels, or `spaxels', spread along the central meridian of Neptune, but not too close to the limb to experience significant mixing with space, as shown in Fig. \ref{pixelpos}. Twelve spaxels, roughly equally spaced in latitude, were selected for detailed description, indicated by the white dots in Fig. \ref{pixelpos} and listed in Table \ref{tbl-2}. Cube `3' was chosen for this analysis as it was representative of all the 120s-integrated observations (`3' -- `6'), and the SNR was sufficient that we did not need to co-add multiple cubes. We originally attempted to fit these data using the noise values provided by the pipeline, but found we could not fit to this level of precision, achieving $\chi^2/n_y$ values of $\sim 4$ at best. Hence, we doubled these errors in the fits presented here, where we achieve a minimum $\chi^2/n_y$  of $\sim 1$. We attribute this inability to fit to within the original noise estimate to `forward-modelling' deficiencies in our model, arising from factors ranging from errors including: 1) errors in our gaseous absorption data; 2) systematic deficiencies in our model set up; 3) the smoothing process itself introducing unaccounted-for covariances into the smoothed spectra; and 4) any residual calibration errors. 

To check the flux calibration, Fig. \ref{nepalbedo} compares the measured disc-averaged reflectance spectrum for Cube `3' with the disc-averaged albedo spectrum of \cite{kark94}, showing the generally good agreement ($\sim10$\%). Neptune has a highly dynamic atmosphere and it is not surprising that the disc-averaged albedo spectrum we have measured in 2018 differs slightly from the spectrum of \cite{kark94}, which was observed in 1993. We can see in Fig. \ref{pixelpos} that the reflectance spectrum changes significantly with both position on the disc and local cloud conditions, as can be seen when we compare these spectra with the observed centre-of-disc reflectance spectrum, which is noticeably brighter. Figure \ref{nepalbedo} also compares the MUSE spectra with recently acquired (September 2018) HST/WFC3 observations near the disc centre (HST/OPAL (e.g. Simon et al. 2015) programme; Simon -- private communication). Here we see reasonably good agreement between the VLT/MUSE and HST/WFC3 observations, which together with our reasonable correspondence with the albedo spectrum of \cite{kark94} leads us to conclude that the photometric calibration of our data is reliable and consistent with previous observations.

\section{Radiative-transfer analysis}
The vertical cloud structure and methane abundance were retrieved from these observations using the NEMESIS radiative transfer and retrieval code \citep{irwin08}.  NEMESIS models planetary spectra either by using a line-by-line model, or by using the correlated-k approximation \citep[e.g.][]{lacisoinas91}. Since our methane absorption data is from a band model \citep{kark10} we used the method of correlated-k here, constructing k-tables from the band data using exponential-sum fitting. As with our previous analyses \citep[e.g.][]{irwin11,irwin16},  we used the matrix-operator multiple-scattering model of \cite{plass73} to simulate these reflected sunlight spectra, modelling the atmosphere with 39 levels spaced  between $\sim 10$ and $\sim 0.01$ bar and using either 5 or 9 angles (both upward and downward) in a Gauss-Lobatto zenith-angle quadrature scheme, with the number of required azimuth components in the Fourier decomposition determined from the maximum of the reflected or incident-solar zenith angles. The collision-induced absorption of H$_2$-H$_2$ and H$_2$-He near 825 nm was modelled with the coefficients of \cite{borysow89a,borysow89b,borysow00}\footnote{Fortran programs for calculating the H$_2$-H$_2$ and H$_2$-He coefficients were downloaded from https://www.astro.ku.dk/~aborysow/programs/index.html.}.  In retrieval tests we found that the H$_2$--H$_2$/H$_2$--He CIA coefficients for hydrogen with thermally-equilibriated ortho:para hydrogen ratio were much more consistent with the observed spectra than the `normal' 3:1 ortho:para hydrogen ratio (more appropriate for planets such as Jupiter, where air is upwelled rapidly from deeper, hotter levels) and so we used these coefficients in our retrievals. A similar conclusion was reached by \cite{baines90} and \cite{gautier95}. Rayleigh scattering was also included, where for each gas the contribution was calculated as described in the Appendix. At these wavelengths, however, effects of polarization and Raman scattering were justifiably ignored \citep[e.g.][]{srom05}. To analyse the measured radiance spectra with our radiative-transfer model we used the solar spectrum of \cite{chance10}, which was first smoothed with a triangular line shape of FWHM = 0.002 $\mu$m to make it compatible with our smoothed spectra. To calculate the reflectivity we also needed Neptune's distance from the Sun on the date of observation, which was 29.94 AU.

The reference temperature and abundance profile used in this study is the same as that used by \cite{irwin14} and is based on the  `N' profile determined by Voyager-2 radio-occultation measurements \citep{lindal92}, with He:H$_2$ = 0.177 (15:85) including 0.3\% mole fraction of N$_2$.  This profile was determined at equatorial latitudes and thus its use as a `global' \textit{a priori} is moot. At the level of the main clouds ($p \sim 2$ bar) it has for many years been thought that Neptune's atmosphere should become barotropic, in which case we would not expect large latitudinal variations in temperature at these levels. However, \cite{tollefson18} have suggested that a vertical shear seen in the zonal wind speeds inferred from tracking cloud features in H-band (1.4--1.8 $\mu$m) and K-band (1.8--2.2 $\mu$m) observations, may be caused by a significant variation in atmospheric temperatures from equator to mid-latitudes to poles. Depending on their assumptions for the latitudinal variation of methane, \cite{tollefson18} find that latitudinal variations in temperature of up to 2 -- 15 K may be present at pressures greater than 1 bar. We will return to the potential effect of such temperature variations on our methane retrievals later in this section. Finally, our \textit{a priori} methane profile was set to have a deep mole fraction of 4\% \citep{kark11}, limited to 100\% relative humidity in the upper troposphere and constrained from exceeding a mole fraction of $1.5\times 10^{-3}$  \citep{lellouch10}  in the stratosphere.

Rather than fit the entire spectrum we focussed only on the 0.77 -- 0.933 $\mu$m region. This was done for a number of reasons: 1) the number of wavelengths was reduced from 459 to 161, which increased retrieval speeds; 2) by limiting the wavelength range to the single reflectance peak containing the H$_2$--H$_2$ CIA feature we could more reasonably assume single particle size distributions; and 3) by analysing longer wavelengths only we are much less sensitive to the effects of Raman scattering, Rayleigh scattering (from air and  stratospheric aerosols), and the effects of polarisation \citep{srom05}. 

We initially attempted to model the spectrum using a continuous profile of identical particles, and found good fits to individual spectra with either small (0.1 $\mu$m) or larger (1.0 $\mu$m) particles. However, we found this model was not able to match the observed limb-darkening/limb-brightening seen at different wavelengths very well and so we instead adopted the parameterised model used by \cite{irwin16} to model VLT/SINFONI and Gemini/NIFS H-band observations of Neptune,  which was found to provide good limb-darkening/limb-brightening behaviour. In this model, particles in the troposphere are modelled with a cloud with a variable base pressure (set to an \textit{a priori} pressure of $\sim 3$ -- 4.5 bar) and a scale height retrieved as a fraction of the pressure scale height, while scattering from haze particles is modelled with a second layer, with variable base pressure at $\sim 0.1$ bar and fixed fractional scale height of 0.1. The particles in the tropospheric cloud were modelled here with a standard Gamma size distribution with mean radius 1.0 $\mu$m and variance 0.05, while the stratospheric haze particles were modelled with a Gamma size distribution with mean radius 0.1 $\mu$m and the same variance of 0.05. Following \cite{irwin15}, the real refractive index of both particle types was set to 1.4 \citep[broadly typical refractive index of H$_2$S and NH$_3$ liquid/ice: ][]{havriliak55,wong74,martonchik94} at a reference wavelength of 0.8 $\mu$m and NEMESIS used to retrieve the imaginary refractive index spectrum. The \textit{a priori} imaginary refractive index spectrum was set to $0.01\pm0.005$ and sampled at every 0.05 $\mu$m between 0.75 and 0.95 $\mu$m, with a `correlation length' of 0.1 $\mu$m set in the covariance matrix to ensure that the retrieved spectrum varied reasonably smoothly with wavelength. At each iteration of the model, the real part of the particles' refractive index spectrum at wavelengths other than the reference wavelength of 0.8 $\mu$m was computed using the Kramers-Kronig relation \citep{sheik05}. Self-consistent scattering properties were then calculated using Mie theory, but the Mie-calculated phase functions were approximated with combined Henyey-Greenstein functions at each wavelength to smooth over features peculiar to perfectly spherical scatterers such as the `rainbow' and `glory', as justified by \cite{irwin18} (although such features are not actually present for the particle sizes we have assumed at these wavelengths in our Mie-modelled phase functions).  We found that this two-cloud-component model was able to adequately represent the limb-darkening/limb-brightening of Neptune's atmosphere at different wavelengths and so would not confuse such effects with the latitudinal changes in our retrieved composition and cloud estimates, since we observe higher latitudes at higher emission angles. 

To set up a baseline model able to account correctly for limb-darkening/limb-brightening we first of all selected observations from Cube `3' in the latitude band $10^\circ$S -- $5^\circ$S, plotted these as a function of observing zenith angle and extracted the general limb-darkening/limb-brightening curves at all wavelengths. At each wavelength, the observations were  fitted with smooth reflectance versus emission angle curves, and sampled at six emission zenith angles between 12.5 and $65^\circ$ (Fig. \ref{limbspectrum}), corresponding to six of the zenith angles in our nine-zenith-angle quadrature scheme, and not coming too near the disc edge to minimise PSF uncertainty errors. In our retrieval we fitted for: the base pressures and opacities of both cloud/haze layers, the fractional scale height of the cloud, the imaginary refractive index spectra of both particle types and a scaling factor for the deep methane abundance. The methane profile was limited to not exceed 100\% relative humidity and to not exceed a mole fraction of $1.5\times 10^{-3}$  in the stratosphere. This gave 16 freely variable parameters in all.  We should stress here that although in this scheme we retrieve a `deep' methane mole fraction, what we are actually sensitive to (and what we actually report) is the abundance of methane just above the main cloud deck, i.e. at 2 -- 3 bar, rather than the `bulk' deep abundance. It is not plausible for the deep methane mixing ratio on Neptune to have a large latitude dependence as this would result in a large gradient in density if not compensated for by a horizontal thermal gradient. The thermal wind equation is actually a density wind equation and if molecular weight gradients existed to great depths then there would also be vertical wind shears extending to great depths, which if integrated to the cloud tops would lead to a huge latitude dependence in the cloud-top wind speed, which is not observed. It is thus likely that any retrieved latitudinal variations in the `deep' methane abundance are restricted to the upper troposphere, as suggested by \cite{kark11} for both Neptune and Uranus in their Fig. 10.  The case for a similar depletion profile on Uranus is made by \cite{srom14,srom19}, based both on theoretical grounds and on the improved fit quality obtained for upper tropospheric depletion rather than deep depletion. 

Our fits to the measured spectra at the six sampled zenith angles are shown in Fig. \ref{limbspectrum} and it can be seen that very good fits were achieved at all zenith angles ($\chi^2/n_y = 1.62$) giving confidence that our model was able to account for limb-darkening/limb-brightening. Our retrieved cloud/haze structure is shown in Fig. \ref{limbretrievals}, together with our retrieved imaginary refractive index spectra. Here we see, as was found in the H-band from VLT/SINFONI observations \citep{irwin16}, that the limb-darkening/limb-brightening is best matched with tropospheric particles with low imaginary refractive index (giving reasonably high single-scattering albedos of $\varpi \sim 0.9$), while in the stratosphere, the particles are retrieved to have higher imaginary refractive index, giving lower single-scattering albedos of $\varpi \sim 0.74$; this higher imaginary refractive index for the haze particles is necessary to lower the single-scattering albedo sufficiently to match the observed limb-darkening at methane-absorbing wavelengths, as previously noted for H-band observations by \cite{irwin11} and \citep{irwin16}. While our model was able to constrain the pressure level of the haze reasonably well, it was less able to constrain the base pressure of the cloud. We thus attempted retrievals with a number of different  base pressures and finally settled on \textit{a priori} pressure of 4.2 bar, which gave the reasonably good fit (i.e. $\chi^2/n_y = 1.62$) to the observations shown in Fig. \ref{limbspectrum}.

Using the assumption that the same cloud scattering properties can be used at all latitudes, we then fixed the complex refractive indices of both particle types, fixed the cloud/haze base pressures and fitted the central meridian spectra with just four parameters: the opacities of the cloud and haze layers, the fractional scale height of the tropospheric cloud, and the deep methane mole fraction scaling parameter. In this analysis we used our 5-zenith-angle quadrature scheme (for reasons of speed) and Fig. \ref{fitspec} shows the fits we were able to achieve to the observed reflectance spectra with our model at our twelve representative locations, spread along the central meridian and summarised in Table \ref{tbl-2}. Here we can see that we achieve a reasonably good fit to the observed spectra with $\chi^2/n_y$ $\sim$ 1--3 at most locations, increasing to larger values between 20 and 40$^\circ$S, where the mid-latitude cloud belt is seen.  Also shown in Fig. \ref{fitspec} are lines joining the observed reflectivities at 827 and 833 nm, extended slightly to enable the gradient to be more clearly seen. These lines will be explained later.

Figure \ref{fitmeth} shows our fitted methane profiles. In most cases we have plotted the retrieved deep methane abundance and errors directly, but at some locations we retrieved very large methane abundances and very large errors. In these locations we found that the pressure level where the retrieved deep methane abundance intersects the saturated methane mole fraction profile lay at pressures greater than the pressure where the cloud optical depth (i.e. opacity to space) is unity. Hence at these locations, since the methane profile is constrained to not exceed 100\% relative humidity, the retrieval becomes insensitive to the deep methane abundance and we can only derive a lower limit on the deep methane abundance, which we have set here to be the saturated mole fraction at the pressure where the cloud optical depth is unity. Apart from these few locations we can see that the deep methane abundances are reasonably well constrained in this analysis and decrease significantly towards the south pole, with the retrieved latitudinal differences mostly exceeding the retrieval errors. As we noted earlier, we used a single temperature profile at all latitudes in this analysis. If the real temperatures were slightly warmer in the  $\sim0.5$ -- 1.5 bar level, the saturated vapour pressure would be higher and thus our retrieved methane profiles would be constrained by different saturated mole fraction profiles. As a result, we would retrieve different deep mole fractions to achieve the same column amount of methane above the cloud top. To test this we computed the saturated methane mole fraction profiles using the latitude-dependent temperature profiles fitted from mid-infrared observations by \cite{fletcher14}, which we also show in Fig. \ref{fitmeth}. Here, we see that since the temperature profiles retrieved by \cite{fletcher14} are generally cooler than our \textit{a priori} assumptions the saturated methane mole fractions are even smaller. Hence, to achieve the same column abundance of methane above the clouds, we would have to either lower the cloud top altitude, or increase further the deep methane abundance. However, mid-IR observations are not very sensitive to the temperatures near the 1-bar level and so, even though \cite{fletcher14} retrieve cooler tropopause/upper-troposphere temperatures, we believe that the cooler temperatures inferred near 1 bar are most likely to be caused more by the vertical smoothing imposed on the profiles, rather than true lowering of the temperatures here although, as we noted earlier, \cite{tollefson18} find that the temperatures in the 1 -- 3 bar region may vary with latitude by several K, depending on the assumed latitudinal variation of methane. If this is the case, then this would have an impact on our retrieved deep mole fractions of methane. Hence, although we detect a clear decrease in methane abundance above the cloud as we go from the equator to the south pole, the absolute abundances are more difficult to constrain.

Fig. \ref{latvar} shows the latitudinal dependence of the retrieved cloud and methane profiles at all 64 `spaxels' analysed along the central meridian as contour plots, where we have plotted the results for all sampled latitudes, not just the twelve shown in Figs. \ref{fitspec} and \ref{fitmeth}.  Fig. \ref{latvar} also shows the fitted $\chi^2/n_y$ at all latitudes together with the latitudinal variation in the retrieved deep abundance of methane. We can see that we achieve good fits to these data at all locations and that we detect a clear and significant decrease in cloud-top methane abundance from 4 -- 5\% at near-equatorial latitudes to 3 -- 4\% polewards of $\sim 40^\circ$S. In addition, Fig. \ref{latvar} also shows the methane latitudinal variation determined from HST/STIS observations on 2003 by \cite{kark11}, showing reasonable correspondence. This indicates that our results are broadly consistent with this earlier determination.

\section{Reflectivity spectral analysis}
To understand how the abundance of methane can be extracted from these data, Fig. \ref{compare_ch4} shows our fit to the spectrum near the disc centre assuming 4\% deep methane mole fraction and then spectra recalculated with the methane abundance either halved or doubled.  It can be seen that as we do this, the modelled reflectivity near the centre of the H$_2$--H$_2$/H$_2$--He CIA bands does not change much, but the reflectivity either side of the 0.820 -- 0.835 $\mu$m reflectance peak is significantly altered. In essence as we increase the abundance of CH$_4$ the peak becomes slightly narrower, especially on the long-wave side.  We wondered if it might be possible to elucidate this phenomenon directly from the observations, without having to run a full and time-consuming retrieval at every location on Neptune's disc. We can see in Fig. \ref{compare_ch4} that at wavelengths 827 and 833 nm there is roughly equal reflectance in our reference spectrum, indicating that we probe reflectance from roughly the same pressure level, but with the opacity of the former wavelength dominated by H$_2$--H$_2$/H$_2$--He CIA absorption and the latter wavelength dominated by CH$_4$ absorption. Hence, as the methane abundance rises, $R_{827} - R_{833}$ increases, and as the methane abundance falls, $R_{827} - R_{833}$ decreases. The observed $R_{827} - R_{833}$ differences are indicated by the short, straight lines shown in Fig. \ref{fitspec}, where we can see that $R_{827} - R_{833}$ is roughly zero near the equator, but becomes noticeably negative as we approach the south pole, consistent with the lower retrieved abundance of CH$_4$ found here.  Fig. \ref{rdiffcube} shows mapped images of the observed reflectivity $R_{827} - R_{833}$  difference applied to all six sets of observations. In this scaling we expect larger reflectivity differences to be associated with enhanced abundances of methane and we can indeed see larger signals near mid-latitudes, falling towards the south pole. We can also see a faint enhancement circling the south pole. As can be seen `striping' artefacts are visible in some of these images, due to calibration errors and there was initially a concern that this might have affected our retrievals. However, it can be seen that the striping is not too severe in observation `3',  chosen for our retrieval analysis, and we could find no significant impact in our retrievals along the central meridian when crossing from one `stripe' to the next. Hence, we conclude that while these small calibration errors are quite noticeable in these radiance-differencing images, which compare two wavelengths, they do not significantly affect our retrievals that simultaneously fit to multiple wavelengths.

In Fig. \ref{compare_image} we show the averaged reflectivity image at 827 nm, averaged over observations `3', `4', and `6', compared with the averaged image at 833 nm and then the averaged difference image. We again see larger $R_{827} - R_{833}$ signals near mid latitudes, falling towards the south pole, and also a slight brightening of the difference at $60 - 70^\circ$S, indicating possible local enhancement of methane at this latitude. Looking at the reflectivity images (which we have here enhanced to try to bring out the weaker features) we see a slightly brighter zone at $60 - 70^\circ$S and then lower reflection polewards of $75^\circ$S in the 827 nm image, whereas the methane image at 833 nm is relatively featureless. Hence, we appear to detect either a slight thickening or vertical extension of the main cloud from $60 - 70^\circ$S which is masked in the methane image by an accompanying increase in methane abundance and thus absorption. We note that an enhancement in methane abundance at $60 - 70^\circ$S can be discerned in our retrieved deep CH$_4$ abundances in Fig. \ref{latvar}.

\section{Principal Component Analysis} The weak and rather noisy nature of the CH$_4$ reflectivity difference map (Fig. \ref{rdiffcube})  led us to explore alternative methods of mapping the spectral relectivity signal of CH$_4$ in Neptune's atmosphere. We turned to the technique of Principal Component Analysis (PCA) \citep[e.g.][]{murtagh87}, used with great success in modelling visible/near-IR Jovian spectra by \cite{dyudina01} and \cite{irwin02}. The basic principle of Principal Component Analysis is that the variance of a set of observed spectra, in this case the varying spectra observed over Neptune's disc, can be decomposed into a set of Empirical Orthogonal Functions (EOFs) that form a linear basis from which any spectrum in the set, $y(\lambda)$, can be reconstructed from a linear combination of the EOFs, $E_i(\lambda)$, as $y(\lambda)=\Sigma_i{\alpha_i E_i(\lambda)}$, where the coefficients, $\alpha_i$, describe the relative proportions of the different EOFs in the combined spectrum. The derived EOFs have with them an eigenvalue, $e_i$, and the EOFs are usually ranked in order of decreasing $e_i$. With this ordering it is  found that most of the variance can be accounted for by the first EOF (i.e. the one with the largest derived eigenvalue), with decreasingly significant  contributions from higher EOFs. The derived EOFs do not necessarily correspond to anything physically significant, but under certain circumstances they can sometimes correspond to real variables.

In this case, since we are interested in searching for the spectral signatures of H$_2$--H$_2$/H$_2$--He and methane absorption, which have similar absorption strengths near 0.83 $\mu$m, we analysed the observed Neptune spectra covering the wavelength range 0.80 -- 0.86 $\mu$m. The results are shown in Fig. \ref{compare_pca}. As can be seen the eigenvalues of the fitted EOFs fall rapidly with increasing EOF number and we can also see that the spatial distribution of the fitted weighting coefficients, $\alpha_i$, become increasingly noisy. In fact, we found that the first three EOFs effectively encapsulate all the significant information. We can see that EOF 1 is mostly  flat, and that its spatial map corresponds almost exactly with the I/F appearance of Neptune over these wavelengths. Hence, this EOF appears to encapsulate the overall observed reflectivity variation.  EOF 2 has a similar spectrum (albeit inverted) and its spatial distribution is almost the inverse of the spatial distribution for EOF 1, although its spatial structure exhibits finer structure, including a dark `halo' about the south pole at $\sim 80^\circ$S. EOF 3 also contains significant spectral variation, but its spatial distribution is very different from that of EOFs 1 and 2, with significant contribution from Neptune's equatorial latitudes, but low contributions towards the south pole. The spectral shape of EOF 3 looks similar to the expected spectral signature of changing the CH$_4$ scaling factor, shown earlier in Fig. \ref{compare_ch4} and reproduced in Fig. \ref{compare_pca}  for reference. Hence, this EOF seems to correspond rather well with the retrieved cloud-top methane abundance and we can see low values polewards of 40$^\circ$S, just as we see in our retrieved methane profies (Fig. \ref{latvar}) and faintly in our reflectivity difference maps (Figs. \ref{rdiffcube} and \ref{compare_image}). In Fig. \ref{latvar} we compare our retrieved methane abundances with the variation of $\alpha_3$ along the central meridian, scaled to match the overall variations, and find that it broadly matches the retrieved reduction in cloud-top methane abundance towards the south pole.  We can also see from both Figs. \ref{latvar} and \ref{compare_pca} that the contribution from EOF 3  has slightly lower values near the equator,  similar to the very slight reduction in retrieved methane abundance near the equator seen in Fig. \ref{latvar}. Hence, applying the PCA technique in this spectral range seems to recover one component that roughly matches the retrieved latitudinal variation of methane cloud-top abundance and which adds credibility to our conclusion that methane at Neptune's cloud tops is diminished at latitudes polewards of 40$^\circ$S at all longitudes and not just along the central meridian as we have formally retrieved.

\section{Conclusion}

We have made an initial estimation of the latitudinal variation of cloud-top methane abundance in Neptune's atmosphere, for the first time from ground-based observations, using the new Narrow Field Mode of the MUSE instrument at VLT.  We find that methane varies with latitude broadly similarly to the variation seen in HST/STIS observations in 2003 \citep{kark11}. Hence, this distribution with latitude would appear to be a long-lived feature (or is at least slowly-varying) and our retrieved abundances appear to be consistent with this earlier determination, with a retrieved mole fraction of  $4 - 5$\% at equatorial latitudes, reducing to $3 - 4$\% at polar latitudes. Our maps of CH$_4$ abundance variations show considerable fine detail, which could be useful for constraining the detailed structure of Neptune's atmospheric dynamics. The lower values of cloud-top CH$_4$ over Neptune's south pole is consistent both with the HST/STIS determinations of \cite{kark11} and also with a multi-wavelength analysis of \cite{depater14}, which concludes that the increased brightness of Neptune's south pole at radio wavelengths is caused by a lower humidity of hydrogen sulphide (H$_2$S). This low humidity may be part of a wider overturning circulation in Neptune's atmosphere, which has air rising at mid-latitudes (detected by cooler tropopause temperatures observed in the mid-IR, \cite[e.g.][]{fletcher14}) and leads to subsiding motion of `dry' air at polar latitudes, extending all the way from the stratosphere down to the cloud tops. In our observations, this upwelling branch is coincident with the cloudy regions seen at 20 -- 40$^\circ$, but what happens at equatorial latitudes is less clear. However, a slightly lower methane abundance apparent at the equator in our retrieved parameters (Fig. \ref{latvar}) and also our EOF 3 contribution map (Fig.\ref{compare_pca}), might indicate local downwelling at the equator also, which may explain why the main cloud here is retrieved to have lower opacity and is less vertically extended. Since we retrieve similar cloud-top pressures at all latitudes, but very different methane abundances, we surmise that this cloud is not composed of methane ice, but is most likely composed of hydrogen sulphide ice, as is probably also the case in Uranus's atmosphere \citep{irwin18, irwin19}.  Interestingly, \cite{depater14} note that South Polar Features (SPFs) seen in Neptune's atmosphere at $60-70^\circ$, are convective storms, produced by baroclinic instabilities at the edge of the south polar prograde jet. The fact that we detect very slightly enhanced methane at these latitudes appears to support this conclusion. Further analysis, combining these VLT/MUSE observations with existing longer-wavelength observations from instruments such as VLT/SINFONI and Gemini/MUSE, is planned to more tightly constrain the cloud properties (using a wider wavelength range) and also explore alternative methods of parameterising the methane vertical profile such as the  ``proportionally descended gas" profiles,  found by \cite{srom11} to provide a good match to the HST/STIS 2002 observations of Uranus. However, this extended analysis is beyond the scope of this `initial-results' paper, reporting the new MUSE/NFM Neptune observations.

\section{Acknowledgements}

We are grateful to the United Kingdom Science and Technology Facilities Council for funding this research. We thank Larry Sromovsky for providing the code used to generate our Rayleigh-scattering opacities. Glenn Orton was supported by NASA funding to the Jet Propulsion Laboratory, California Institute of Technology. Leigh Fletcher was supported by a Royal Society Research Fellowship at the University of Leicester. PMW received support from BMBF Verbundforschung (project MUSE-NFM, grant 05A17BAA). The observations reported in this paper have the ESO ID: 60.A-9100(K).



{\it Facilities:}   \facility{VLT (MUSE)}.

\section{Appendix - Calculation of Rayleigh-scattering optical depth}

The Rayleigh-scattering cross-section of Neptune's air (i.e. cm$^{2}$) in our radiative transfer model is calculated as \citep[e.g.][]{allen76}:

$$ \sigma(\lambda) = {8 \pi^3 \over {3 \lambda^4 N^2}}  {\sum_i{x_i(n_i(\lambda)^2-1)^2 \delta_i} \over {\sum_i x_i}}$$

where the summation is over each gas present, with mole fraction $x_i$, refractive index spectrum $n_i(\lambda)$ (at standard temperature and pressure) and depolarization factor $\delta_i$. $N$ is the total number density of air molecules (i.e. cm$^{-3})$. The depolarisation factor for each gas is calculated as:
$$ \delta_i = {{6+3\Delta_i} \over {6-7\Delta_i}} $$

where the factors $\Delta_i$ are listed in Table \ref{tbl-3}. Similarly, the refractive index of each gaseous component (at STP) is taken to vary with wavelength as
$$n_i(\lambda) = 1 + A_i(1+B_i/\lambda^2)$$

The approach of summing over mole fraction like this was suggested by Larry Sromovsky [private communication], who also shared the parameters used. The coefficients  $A_i$ and $B_i$ are also listed in Table \ref{tbl-3}.  Here, the refractive index parameters are taken from \cite{allen76}, although for CH$_4$, which is not listed, they have been set to  be the same as NH$_3$. The depolarization coefficients, $\Delta_i$ are from  \cite{penndorf57}, after \cite{parthasarathy51}; CO$_2$ values have been assumed for CH$_4$ and NH$_3$.

\clearpage



\begin{figure}
\epsscale{0.8}
\plotone{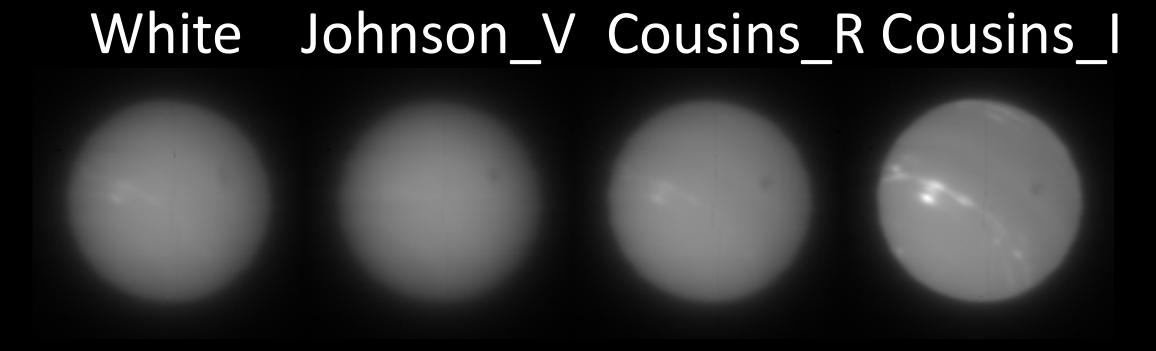}
\caption{VLT/MUSE observation `3' analysed in this study averaged over all wavelengths (i.e. white) and also over three different photometric spectral channels: a) Johnson V (507 -- 595 nm); b) Cousins R (578 -- 716 nm); and c) Cousins I (712 -- 861 nm). In these plots, Neptune's south pole is at bottom left. It can be seen that deeper cloud structure becomes increasingly clear at longer wavelengths, where we are less sensitive to Rayleigh scattering from overlying gas molecules.\label{obs_image}}
\end{figure}

\begin{figure}
\epsscale{0.8}
\plotone{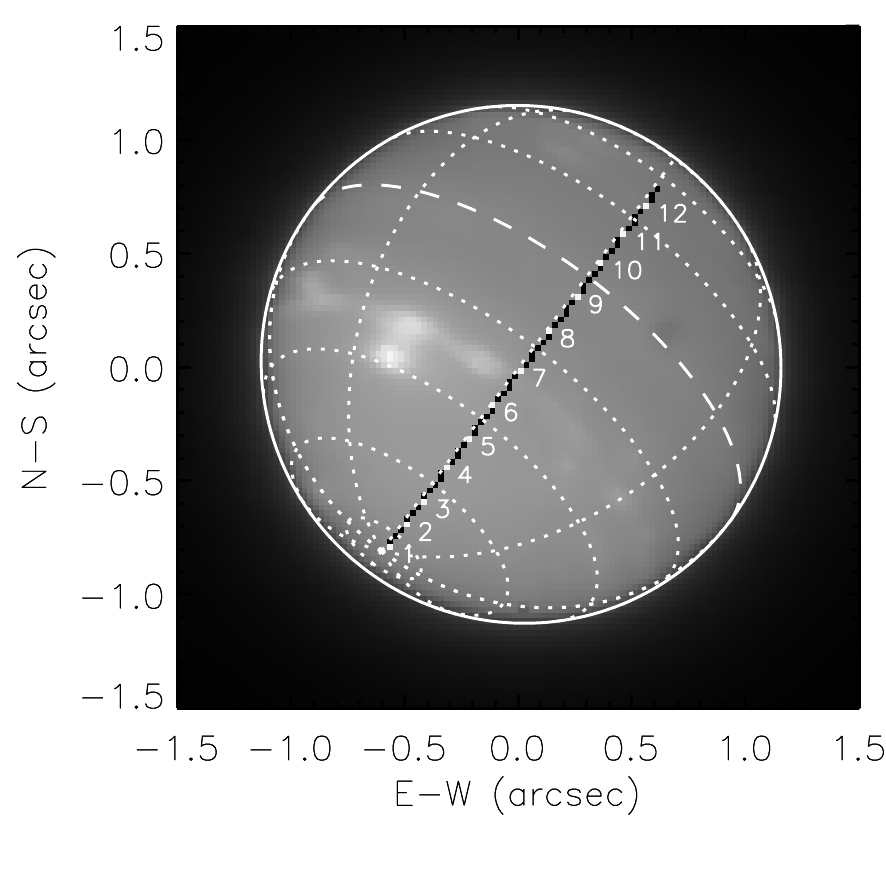}
\caption{VLT/MUSE observation `3' averaged over the Cousins I photometric channel (712 -- 861 nm), plotted as apparent size and angle on the sky, and showing the position of the sampled `spaxels' along the central meridian (black squares). The spaxels selected for detailed description in subsequent plots and tables are indicated by the white squares. Also plotted is a reference latitude-longitude grid, with lines of latitude (planetocentric) separated by 20$^\circ$ and lines of longitude separated by 45$^\circ$. The equator is indicated by the dashed line, rather than dotted line, and the south pole is at bottom left. \label{pixelpos}}
\end{figure}

\begin{figure}
\epsscale{0.8}
\plotone{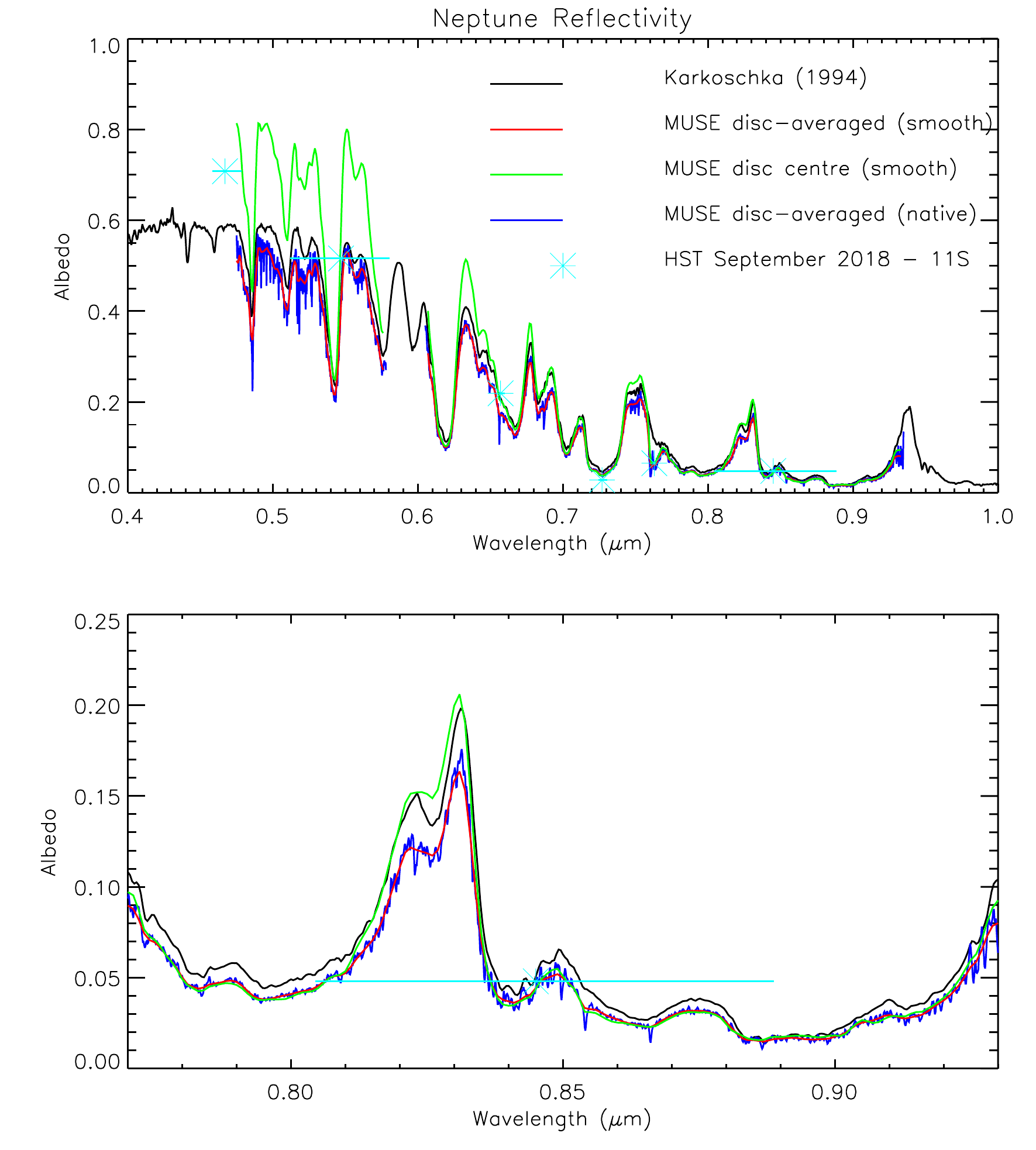}
\caption{Reflectivity spectra of Neptune observed with VLT/MUSE compared with the albedo spectrum of \cite{kark94} and recent observations  with HST/WFC3 (September 2018). The top plot shows the whole VLT/MUSE spectral range, while the bottom plot shows the $0.77-0.93$ $\mu$m region only. Three different MUSE spectra are shown: 1) a disc-averaged spectrum at MUSE's native spectral resolution; 2) a disc-averaged spectrum smoothed to a resolution of 0.002 $\mu$m; and 3) a measured, smoothed MUSE spectrum at the centre of Neptune's disc. Recent photometic observations by HST/WFC3 (September 2018) are shown in cyan for comparison, with the approximate width of the filter channels indicated by the horizontal lines.\label{nepalbedo}}
\end{figure}

\begin{figure}
\epsscale{0.4}
\plotone{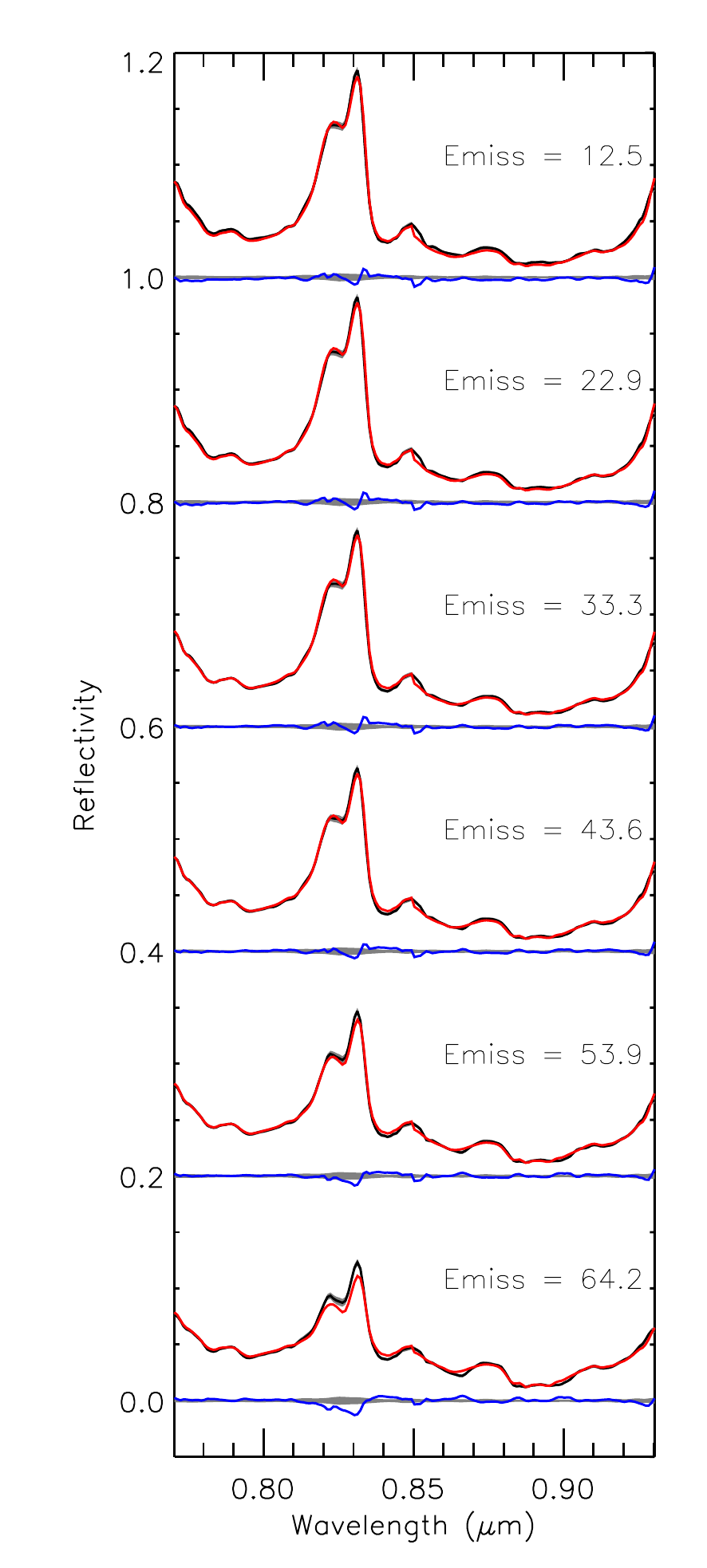}
\caption{Fit to the average observed I/F spectra in Observation `3' in the 5 -- 10$^\circ$S latitude band sampled at six viewing zenith angles corresponding with angles in the nine-zenith-angle Gauss-Lobatto quadrature scheme used in our analysis. Spectra are plotted with same vertical scale but successively offset by a reflectivity difference of 0.2 for clarity. The black lines are the observed spectra (with errors indicated in grey) while our fits are shown in red. At each angle, the difference is shown below in blue, together with the estimated measurement errors, again in grey.\label{limbspectrum}}
\end{figure}

\begin{figure}
\epsscale{0.6}
\plotone{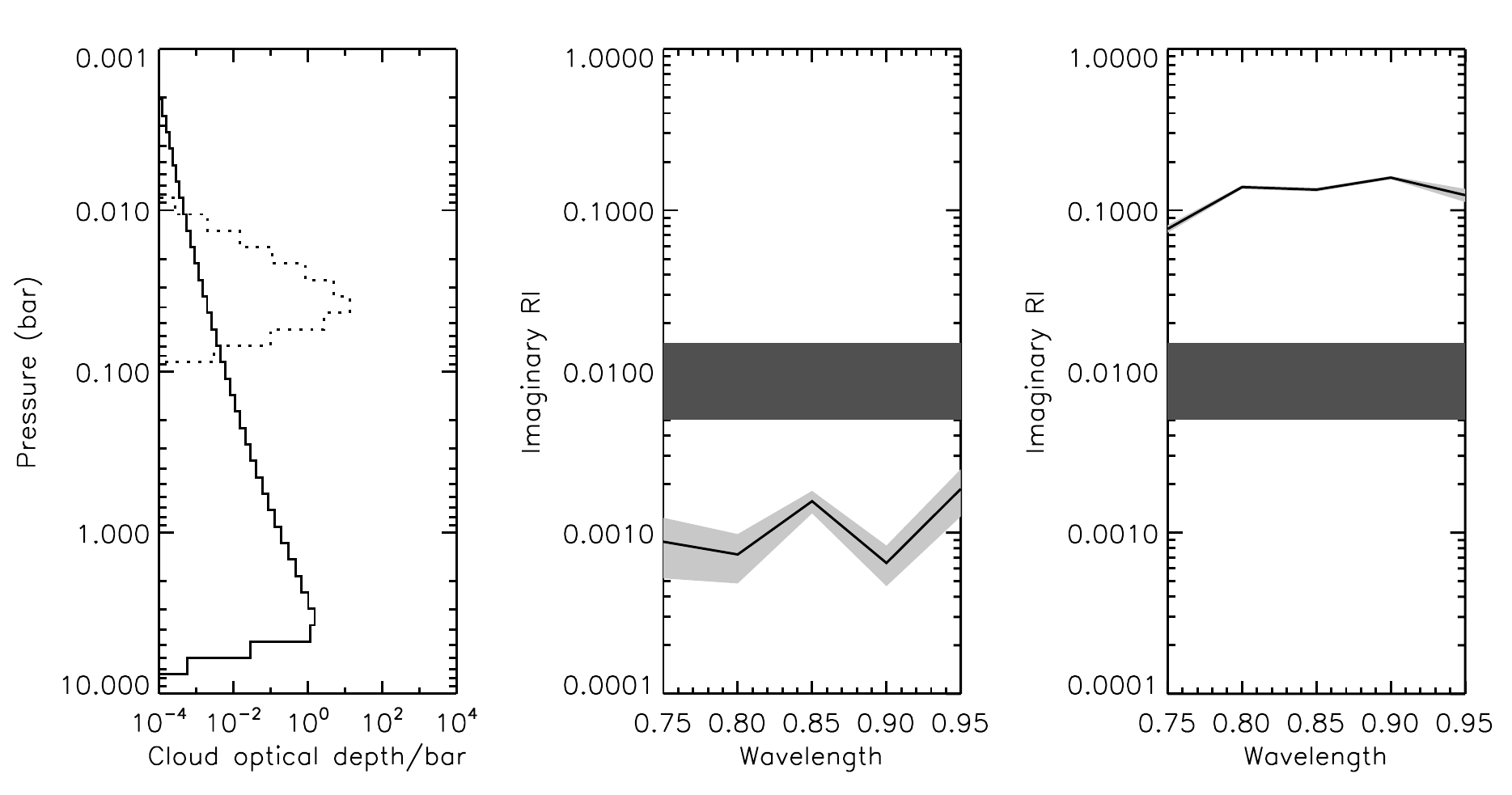}
\caption{Retrieved cloud structure and imaginary refractive index spectra from our limb-brightening/limb-darkening analysis. The left hand panel shows our fitted cloud/aerosol vertical profiles (in opacity/bar at 0.8 $\mu$m), with the vertical distribution of 1.0-$\mu$m cloud particles shown as the solid curve, and the distribution of the 0.1-$\mu$m haze particles shown as the dotted-line curve. The middle and right hand panels show the retrieved imaginary refractive index spectra of the cloud and haze particles, respectively. Here the retrieved spectra and errors are shown by the solid lines and light grey regions. The dark grey regions show the assumed \textit{a priori} imaginary refractive index range. Here, we see that the cloud particles are retrieved to have low imaginary refractive index (and are thus quite scattering), while the haze particles are retrieved to have high imaginary refractive index (and are thus quite absorbing). \label{limbretrievals}}
\end{figure}

\begin{figure}
\epsscale{0.7}
\plotone{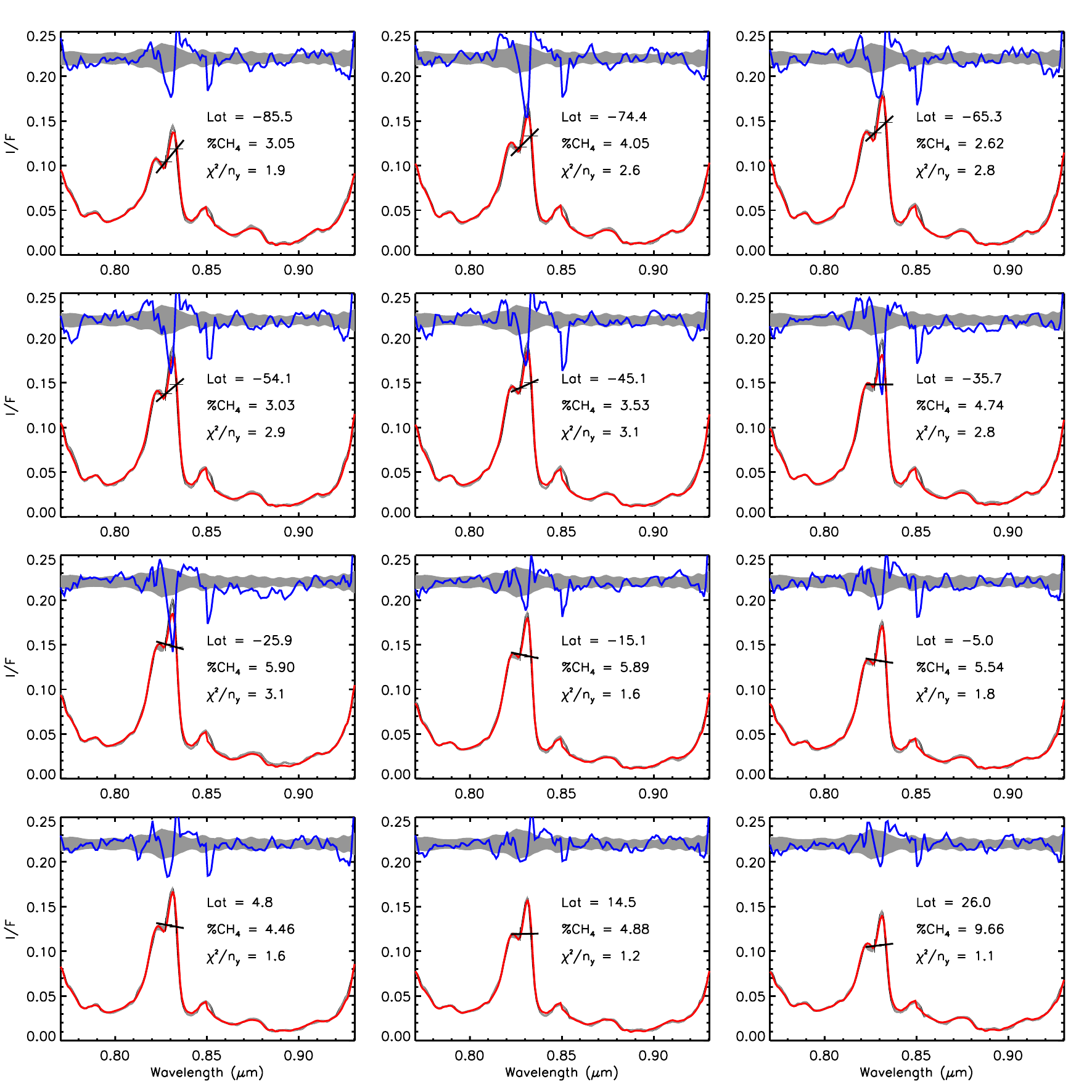}
\caption{Fits to the observed spectra at our twelve sample locations (summarised in Table \ref{tbl-2}). Here the observed spectra, including error bars, are shown in grey, while our fitted spectra are shown in red. At the top of each plot we also show the difference between the measured and fitted spectra (in blue), multiplied by 5 and offset for clarity, together with the error range in grey (also multiplied by 5). It can be seen that we achieve excellent fits with $\chi^2/n$ of the order of 1.0 -- 3.0. The near-horizontal lines near the reflectance peak link the observed reflectivities at our chosen test wavelengths of 827 and 833 nm, as explained in the text. Positive slopes indicate less methane, negative slopes indicate more methane. \label{fitspec}}
\end{figure}

\begin{figure}
\epsscale{0.5}
\plotone{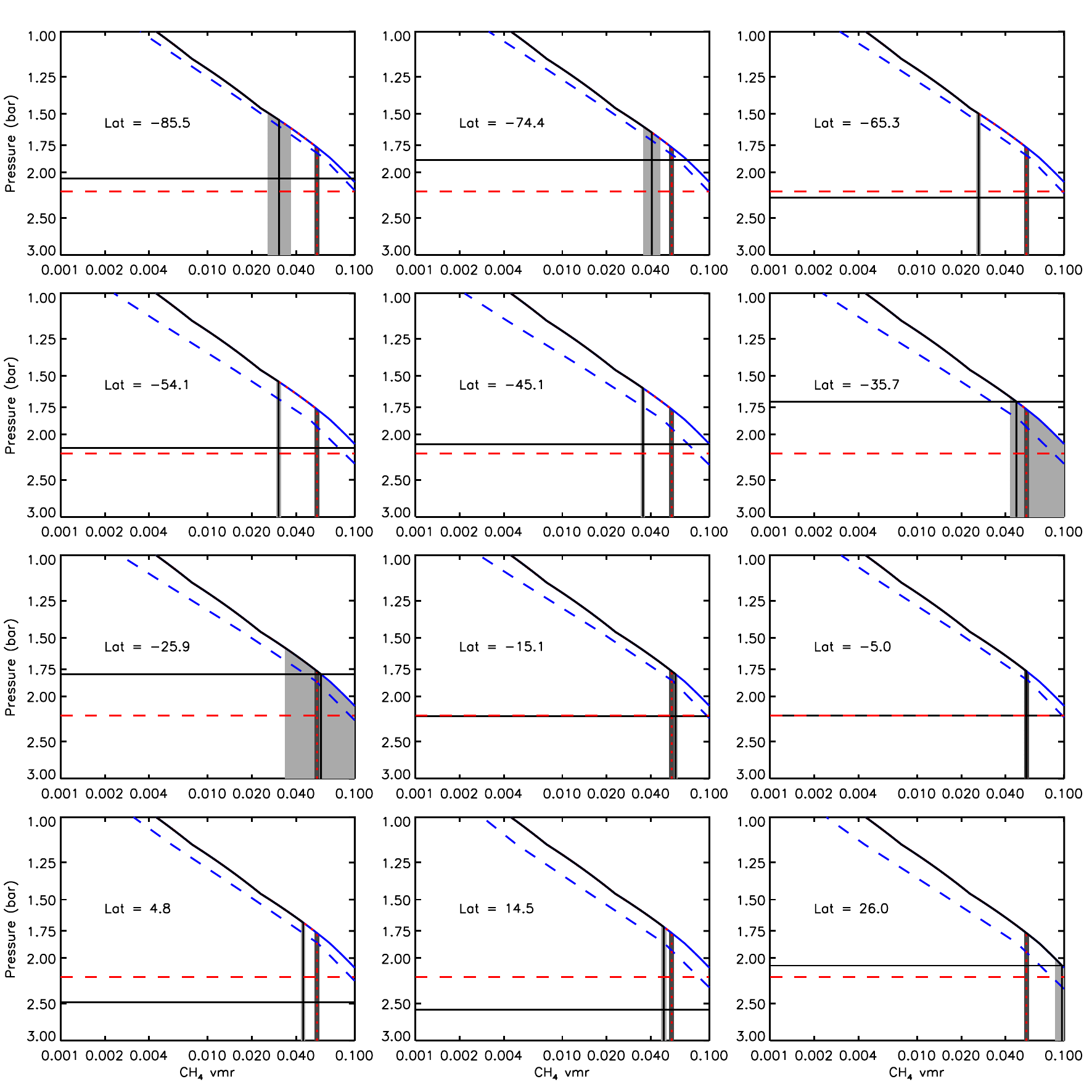}
\caption{Fitted methane mole fraction profiles at our sample locations, summarised in Table \ref{tbl-2}.  Our methane profile is parameterised to have a constant  mole fraction at deep pressures limited to not exceed 100\% relative humidity at lower pressures,  with the saturated mole fraction calculated from our assumed reference temperature profile and shown here as the blue line. To aid comparison, the methane profile retrieved for the reference pixel area `9' at $5^\circ$S is shown in red in all panels. In these plots the uncertainty of the profiles (modified as described in the text) are indicated in grey. A darker grey has been used to indicate the profile error for the reference pixel area `9' at 5.1$^\circ$S. We can see that polewards of $35^\circ$S there is significantly less methane than at equatorial latitudes. The dashed blue lines show the saturated methane mole fraction corresponding to the temperature profiles retrieved at these latitudes by \cite{fletcher14}. Since these temperature profiles are generally cooler than the reference temperature profile, the saturated methane mole fractions are generally smaller. The horizontal black lines show the pressure where the cloud optical depth at 0.8 $\mu$m (i.e. opacity to space) is unity, while the horizontal, dashed red line shows this pressure level for the reference pixel area `9'. 
 \label{fitmeth}}
\end{figure}

\begin{figure}
\epsscale{0.45}
\plotone{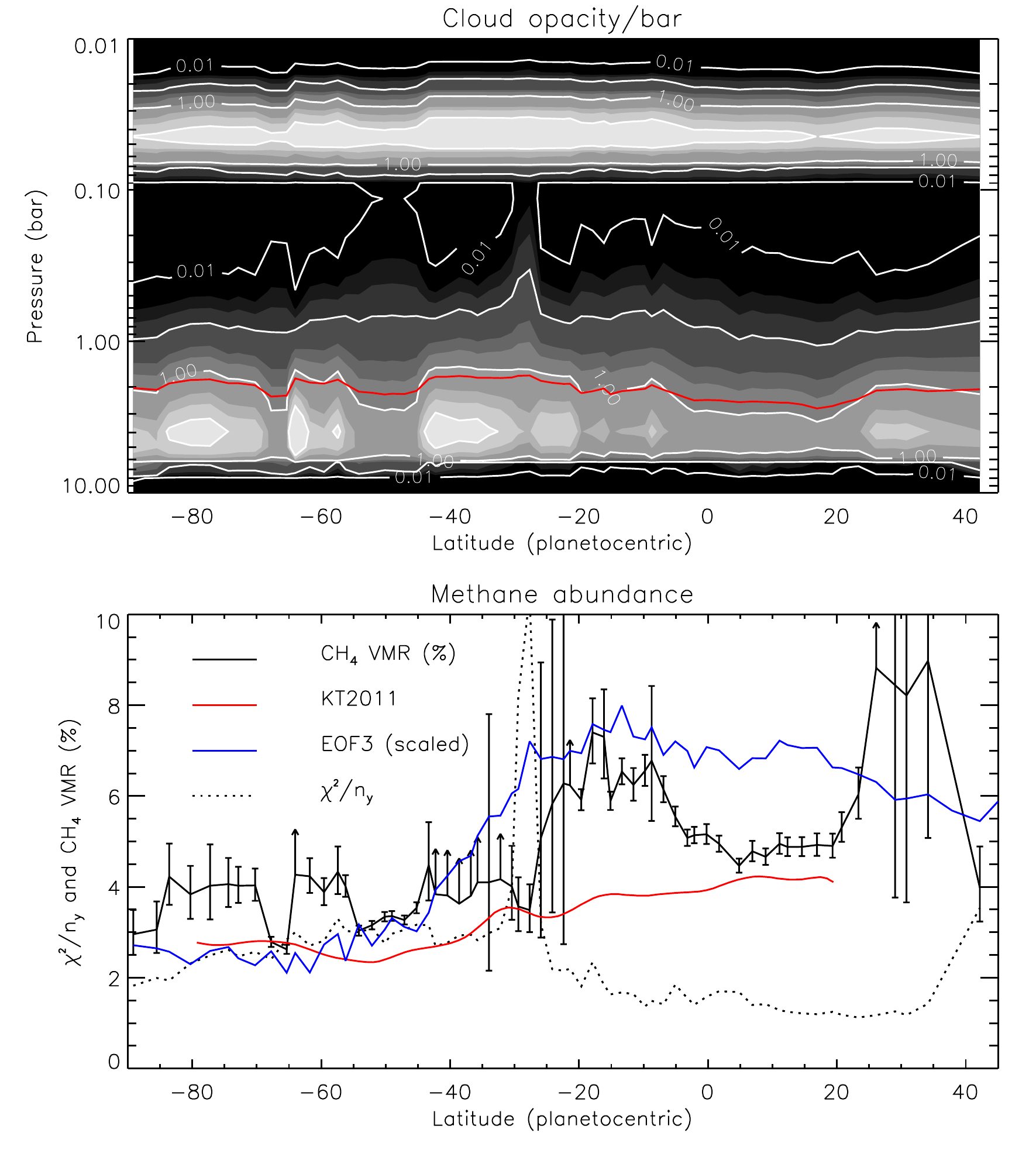}
\caption{Latitudinal dependence of retrieved cloud opacity/bar (at 0.8 $\mu$m), deep methane mole fraction and $\chi^2/n_y$. The top panel shows a contour plot of the fitted cloud opacity/bar profiles at all `spaxels' sampled along the central meridian (brighter regions indicate greater cloud density, with black indicating an opacity/bar of zero and white indicated an opacity/bar of 50). Additional contour lines are drawn for opacity/bar values of 0.01, 0.1, 10 and 50. The bottom panel shows the latitudinal variation of the retrieved deep methane mole fraction, with error bars, and the $\chi^2/n_y$ of the fits. Where the cloud top intersects with the vapour pressure curve of methane, we can only estimate a lower limit of the deeper mole fraction, and indicate this with an arrow symbol instead of an error bar. We see a clear decrease in methane abundance from 5--6\% at near-equatorial latitudes to 3--4\% polewards of 20--40$^\circ$S. The red line in the top plot indicates the cloud top pressure (i.e. level where overlying cloud opacity at 0.8 $\mu$m is unity). The cloud contour map indicates the main cloud top to lie at similar pressure levels at all latitudes and has a cloud-top pressure of $\sim  2$ bar. We can also see an increase in stratospheric haze at 20 -- 40$^\circ$S, associated with the cloudy zone and then clearing towards the north and south. The bottom plot also shows the methane abundance estimated by \cite{kark11} (red), together with the scaled central meridian variation of the contribution of the 3rd Empirical Orthogonal Function we derive with our PCA analysis (blue) in section 5. \label{latvar}}
\end{figure}

\begin{figure}
\epsscale{0.8}
\plotone{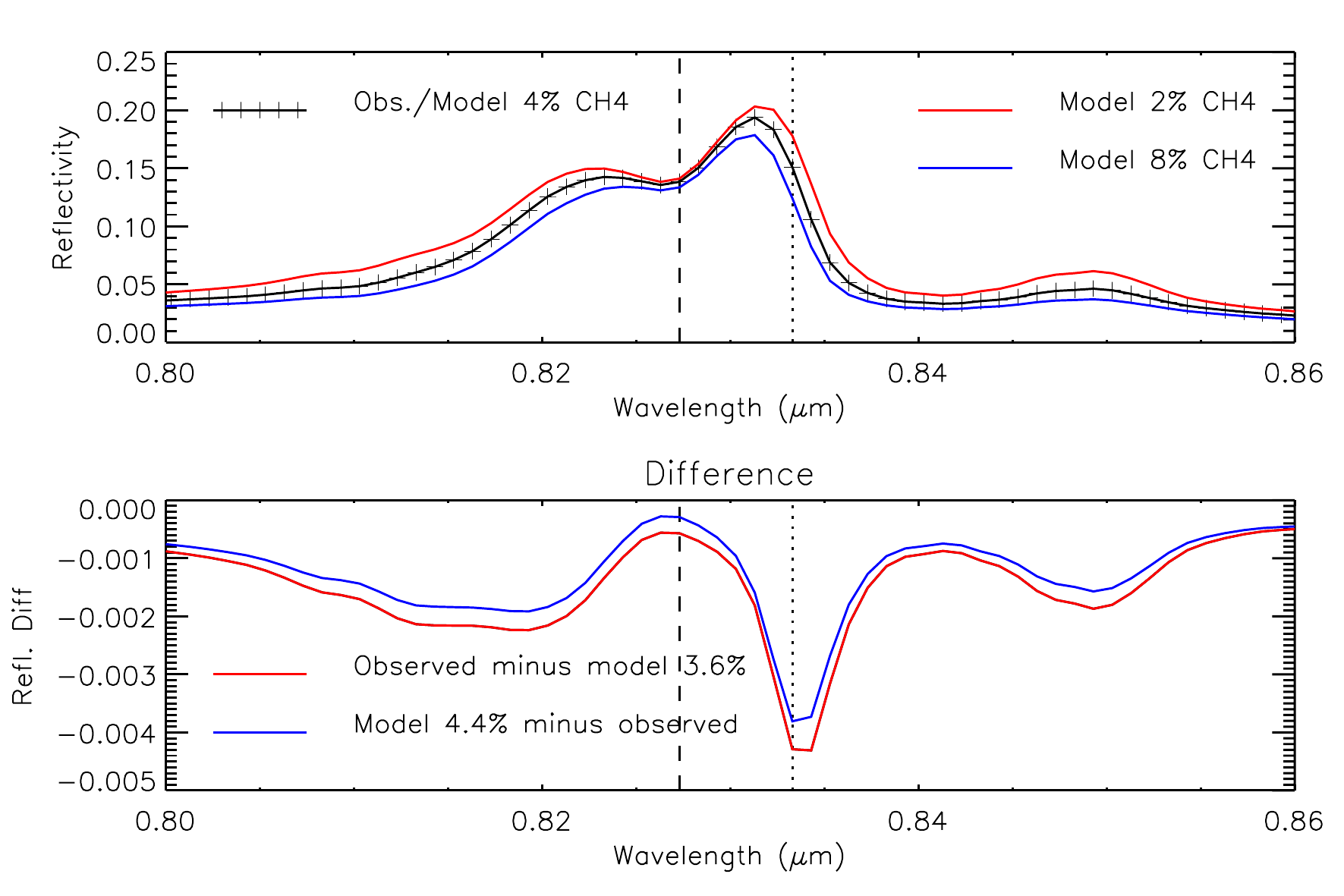}
\caption{Top: Observed MUSE spectrum at disc centre ($\sim 20^\circ$S, cross symbols) near 825 nm and fitted spectrum (black line), where the cloud opacity profiles and cloud imaginary refractive indices have been allowed to vary, but the methane profile has been fixed to the  \textit{a priori} methane abundance profile, which has a deep mole fraction of 4\%. The modelled spectrum recalculated with half or double this deep methane abundance is shown by the red and blue lines, respectively.  Also indicated by the vertical lines are the wavelengths chosen for reflectance differencing, which probe regions of similar reflectivity (and thus similar pressure levels) for the reference methane abundance of 4\%, but whose opacities are predominantly determined by H$_2$--H$_2$/H$_2$--He CIA and methane absorption, respectively. Bottom: change in modelled reflectivity when the methane abundance is increased (blue) or decreased (red) by 10\% from the nominal value of 4\%.\label{compare_ch4}}
\end{figure}

\begin{figure}
\epsscale{0.8}
\plotone{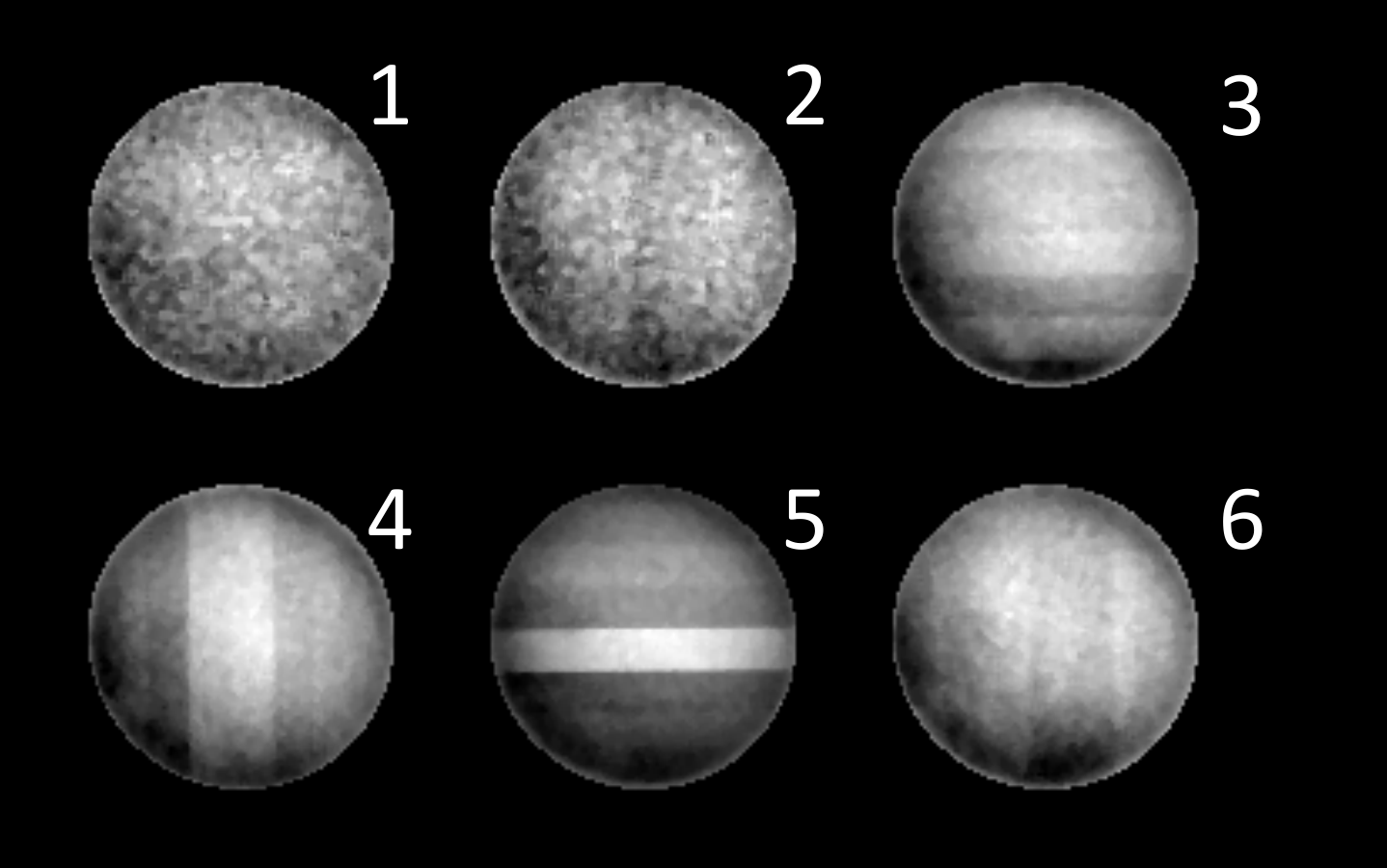}
\caption{Reflectivity differencing method applied to all six individual MUSE observations, indicated by the labels `1' -- `6', showing the reflectivity at 827 nm minus the reflectivity at 833 nm, i.e. $R_{827} - R_{833}$. Here, as in Fig. \ref{pixelpos} the south pole is at bottom left. The noisiness of the first two observations (which were only integrated for 10s, rather than 120s) is clear. For the remaining observations `3' -- `6', these were measured with Neptune centred at different positions within the MUSE FOV and with different position angles (corrected for here). It is clear that residual photometric calibration `striping' artefacts are apparent when comparing just two wavelengths as we do here (i.e. the horizontal and vertical stripes). However, for all these observations, it can be seen that the images are slightly darker at the bottom left than top right, indicated a lower methane spectral signal near the south pole. Approaching the south pole a faint arc (or perhaps even part of a ring) of enhanced $R_{827} - R_{833}$ can just be discerned in cubes `3' and `6', indicating possibly enhanced methane at $60 - 70^\circ$S. \label{rdiffcube}}
\end{figure}

\begin{figure}
\epsscale{0.8}
\plotone{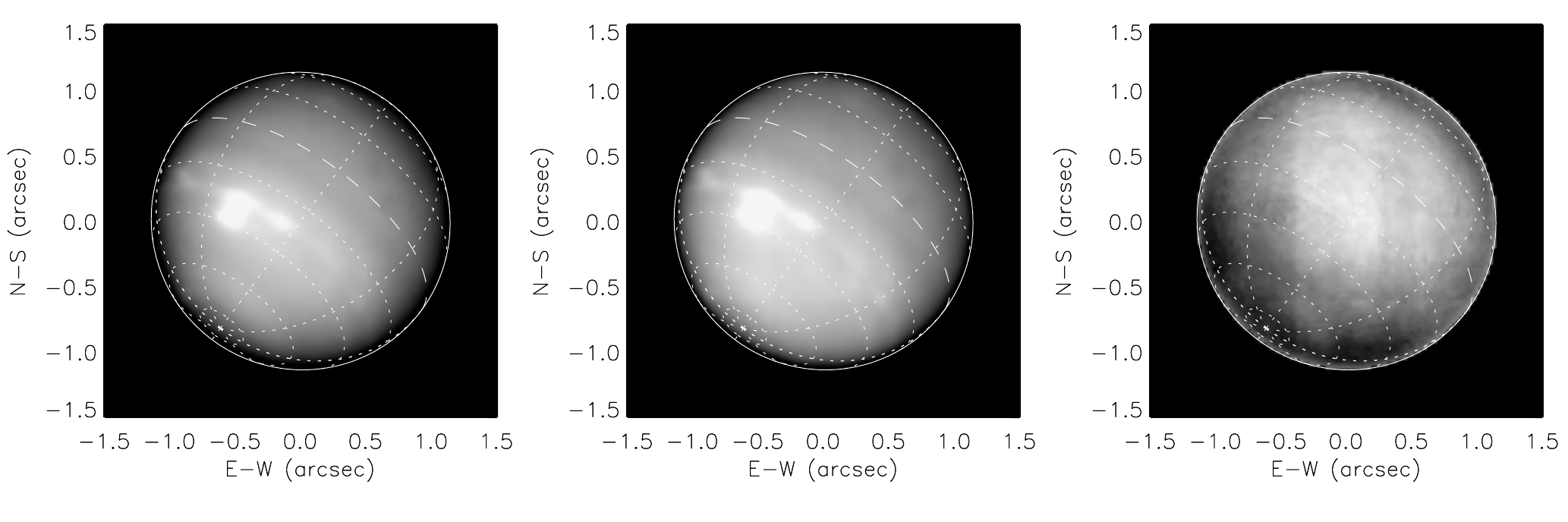}
\caption{Averaged and contrast-enhanced reflectivity observations in the H$_2$--H$_2$/H$_2$--He CIA band at 827 nm (left-hand panel) and methane-band at 833 nm (middle panel) and the difference between these averages (i.e. $R_{827}-R_{833}$) (right-hand panel), all averaging over observations `3', `4', and `6' (observation `5' was omitted as the striping artefacts were too noticeable). The south pole of Neptune is at bottom left, and latitude circles with spacing of 20$^\circ$ are indicated. It can be seen that the reflectivity difference image (right hand panel) is slightly darker at the bottom left than top right, indicated a lower methane spectral signal near the south pole. We can also see a slightly brighter region circling the south pole from $60 - 70^\circ$S, indicating slightly higher methane abundance there, which is also just seen in our retrieved latitudinal dependence (Fig. \ref{latvar}). This feature is just apparent in the H$_2$--H$_2$/H$_2$--He image (left-hand panel), but not the methane-band image (middle panel), indicating that at $60 - 70^\circ$S we seem to have a slight thickening or vertical extension of the main cloud, which is masked in the methane image by an accompanying increase in methane abundance and thus absorption.\label{compare_image}}
\end{figure}

\begin{figure}
\epsscale{0.4}
\plotone{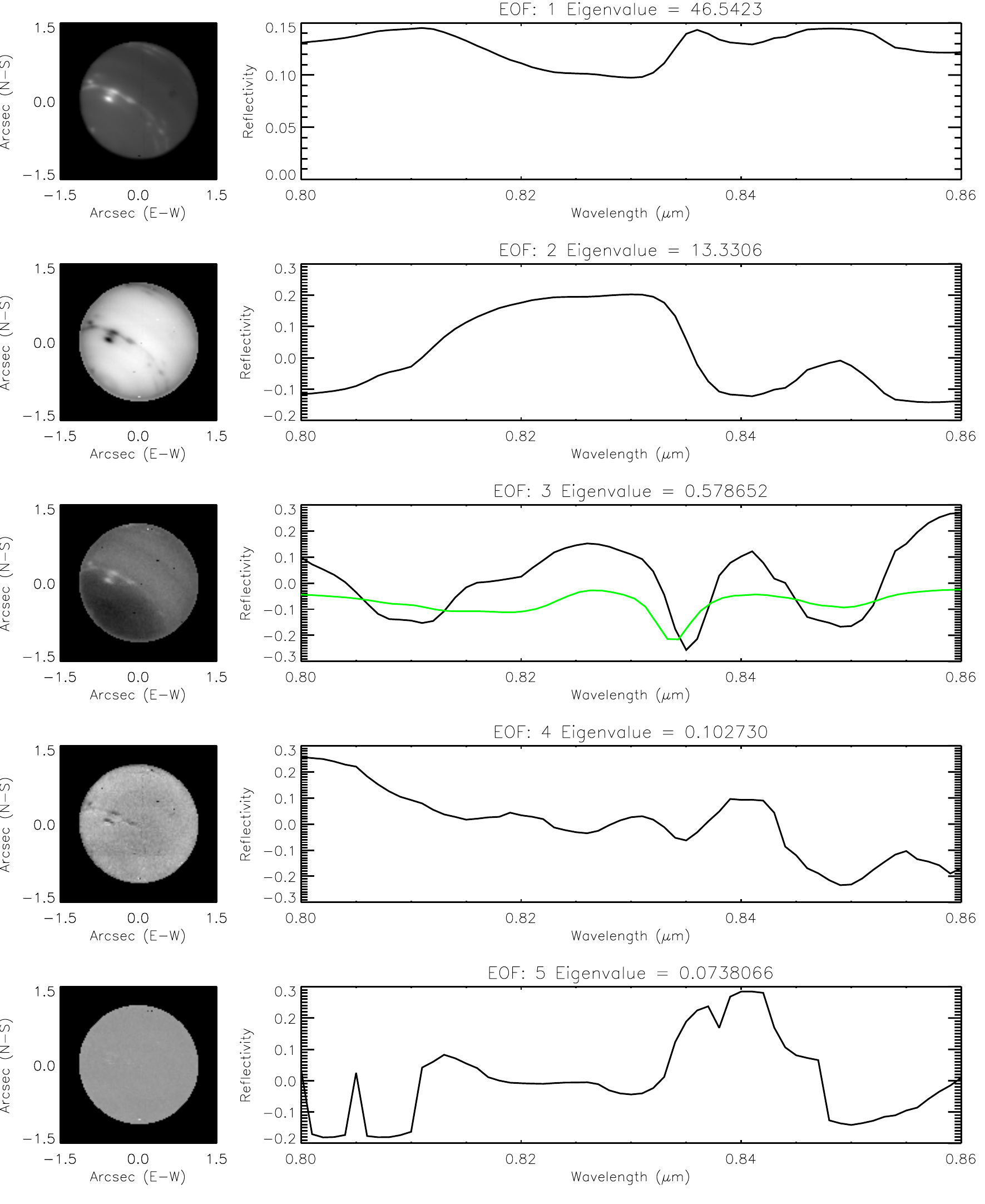}
\caption{Principal component analysis applied to MUSE observation `3', showing the contribution of the first five Empirical Orthogonal Functions (EOF).  The images in the left-hand column show the spatial variation of contribution of each EOF to the reconstructed spectra, while the right hand column shows the associated EOF itself (i.e. spectrum). In the images, Neptune's south pole is again at bottom left.  It can be seen that the eigenvalues of the EOFs fall rapidly with EOF number, and that most meaningful spatial variation in the observation is encapsulated in the first three EOFs. The shape of EOF 1 is approximately flat and this eigenfunction seems to encapsulate the overall reflectivity as can be seen in the associated image. EOF 2 appears to be almost the inverse of EOF 1, both in terms of its spatial distribution and also spectral shape, although its spatial structure exhibits finer structure, including a dark `halo' about the south pole. The spatial distribution of EOF 3 has low values near Neptune's south pole, but high values at lower latitudes, and its spectral shape is rather distinct. The spectral shape of EOF 3 is compared here with the computed change in modelled spectrum when the abundance of CH$_4$ is increased (green), taken from Fig. \ref{compare_ch4}, where we can see a rough correspondence. Hence, we believe EOF 3 can be taken, to a first approximation, as a proxy for the abundance of CH$_4$ immediately above Neptune's main cloud top at $\sim 2$ bar. 
\label{compare_pca}}
\end{figure}

\begin{table}[!h]
\caption{MUSE/NFM Neptune observations.\label{tbl-1}}
\begin {tabular}{l l l l l }
\hline
Obs. ID. & Date & Time (UT) & exposure time & airmass \\
\hline
1 & June 19th 2018 & 09:37:41 & 10s  & 1.0590 \\
2 & June 19th 2018 & 09:40:10 & 10s  & 1.0579 \\
3 & June 19th 2018 & 09:43:21 & 120s & 1.0563 \\
4 & June 19th 2018 & 09:47:44 & 120s & 1.0550 \\
5 & June 19th 2018 & 09:52:06 & 120s & 1.0541 \\
6 & June 19th 2018 & 09:56:26 & 120s & 1.0535 \\
\hline
\end {tabular}

\small {The full-width-half-maximum  (FWHM) of the point spread function (PSF) of these adaptive optics observations is estimated to be 0.06\arcsec at 0.8 $\mu$m. }

\end{table}

\begin{table}[!h]
\caption{Retrieval results at selected areas (12 of 64) considered on Neptune's disc.\label{tbl-2}}
\begin {tabular}{l l l l l l l l l l l}
\hline
Area & Latitude & $p_1$ & $f_{CH_4}$  & $p_C^*$ & $\tau_C$ & FSH$_C$ &
$p_H^*$  & $\tau_H$ & FSH$_H^*$ & $\chi^2/n_y$  \\
 & & (bar) & \% & (bar) & & & (bar) & & & \\
\hline
1 & 85.5$^\circ$S & 2.06 & $3.06 \pm 0.57 $ & 4.3 & 14.13 & 0.312 & 0.038 & 0.20 & 0.1 & 1.99   \\
2 & 74.4$^\circ$S & 1.88 & $4.06 \pm 0.49$ & 4.3 & 24.81 & 0.289 & 0.038 & 0.27 & 0.1 & 2.62  \\
3 & 65.4$^\circ$S & 2.26 & $2.62 \pm 0.57$ & 4.3 & 6.71 & 0.395 & 0.038 & 0.23 & 0.1 & 2.84   \\
4 & 54.2$^\circ$S & 2.14 & $3.03 \pm 0.46$ & 4.3 & 7.40 & 0.413 & 0.038 & 0.29 & 0.1 & 3.00   \\
5 & 45.2$^\circ$S & 2.10 & $3.54 \pm 0.57$ & 4.3 & 7.30 & 0.424 & 0.038 & 0.30 & 0.1 & 3.17   \\
6 & 35.8$^\circ$S & 1.70 & $> 4.74 $ & 4.3 & 33.75 & 0.297 & 0.038 & 0.47 & 0.1 & 2.82  \\
7 & 25.9$^\circ$S & 1.80 & $5.90 \pm 1.07$ & 4.3 & 21.56 & 0.325 & 0.038 & 0.44 & 0.1 & 3.18   \\
8 & 15.1$^\circ$S & 2.20 & $5.89 \pm 1.00$ & 4.3 & 6.91 & 0.399 & 0.038 & 0.31 & 0.1 & 1.63  \\
9 & 5.1$^\circ$S & 2.20 & $5.55 \pm 1.13$ & 4.3 & 6.65 & 0.411 & 0.038 & 0.31 & 0.1 & 1.85\\
10 & 4.8$^\circ$N & 2.49 & $4.46 \pm 1.16$ & 4.3 & 4.95 & 0.392 & 0.038 & 0.23 & 0.1 &1.68  \\
11 & 14.6$^\circ$N & 2.58 & $4.88 \pm 1.04$ & 4.3 & 4.51 & 0.387 & 0.038 & 0.20 & 0.1 & 1.21  \\
12 & 26.1$^\circ$N & 2.07 & $> 9.67$ & 4.3 & 14.51 & 0.305 & 0.038 & 0.24 & 0.1 & 1.17  \\
\hline
\end {tabular}

\small {Notes: variables marked with an asterisk $^*$ were all fixed to values found through our limb-darkening analysis; $p_1$ is the pressure (bar) where the cloud opacity to space is unity; $f_{CH_4}$ is the retrieved cloud-top CH$_4$ mole fraction (\%), i.e. the mole fraction at pressure $p_1$;  $p_C$ is the base pressure (bar) of the lower cloud;  $\tau_C$ is the opacity of the cloud (at 0.8 $\mu$m);  FSH$_C$ is the fractional scale height of the cloud; $p_H$ is the base pressure (bar) of the haze;  $\tau_H$ is the opacity of the haze (at 0.8 $\mu$m);  FSH$_H$ is the fractional scale height of the haze; $\chi^2/n_y$ is the chi-squared statistic of the fit. Where the methane mole fraction at the cloud-top pressure $p_1$ is already limited to not exceed the saturated vapour mole fraction, we instead quote a lower limit for the methane abundance just below the cloud. The assumed refractive index spectra (deduced from the limb-darkening analysis) of the cloud and haze particles are shown in Fig. \ref{limbretrievals}}

\end{table}

\begin{table}[!h]
\caption{Assumed Rayleigh-scattering parameters.\label{tbl-3}}
\begin {tabular}{l l l l}
\hline
Gas & $A_i$ & $B_i$ & $\Delta_i$ \\
\hline
H$_2$ & $13.58 \times 10^{-5}$ & $7.52 \times 10^{-3}$ & 0.0221 \\
He & $3.48 \times 10^{-5}$ & $2.3 \times 10^{-3}$ & 0.025 \\
CH$_4$ & $37 \times 10^{-5}$ & $12 \times 10^{-3}$ & 0.0922 \\
NH$_3$ & $37 \times 10^{-5}$ & $12 \times 10^{-3}$ & 0.0922 \\
\hline
\end {tabular}

\small {Air refractive index calculated as $n_i = 1 + A_i(1+B_i/\lambda^2)$, where $\lambda$ is the wavelength in $\mu$m. $\Delta_i$ terms are the assumed depolarization factors. The literature sources for the coefficients presented here are described in the Appendix. }

\end{table}

\end{document}